\documentclass[a4paper,11pt]{article}
\usepackage{amssymb,amsmath,bm}  % for math
\usepackage{graphics,graphicx}           % for figures
\usepackage{array,booktabs}               % for tables
\usepackage{authblk}                          % for footnote style author/affiliation

\usepackage[margin=4cm]{geometry}
\usepackage{cite}
\usepackage{color}% for various colors
\definecolor{coolblack}{rgb}{0.0, 0.18, 0.39}
\usepackage[linktocpage,colorlinks = true,linkcolor = blue,urlcolor  = coolblack,citecolor = red,anchorcolor = green]{hyperref}

\title{Vacuum and symmetry breaking: new approach}
\author[1,2]{Ahmad Mohamadnejad\thanks{mohamadnejad.a@lu.ac.ir}}
\affil[1]{Department of Physics, Lorestan University, Khorramabad, Iran}
\affil[2]{School of Particles and Accelerators, Institute for Research in Fundamental Sciences (IPM), Tehran, Iran}
\date{\today}

\begin{document}

\baselineskip 0.65 cm

\maketitle

\begin{abstract}
We propose a new mechanism for symmetry breaking in which, apart from particle degrees of freedom, topological degrees of freedom also emerge. In this method, a decomposition for the fields of the Yang-Mills-Higgs theory is introduced and Lagrangian is written based on new variables. This new Lagrangian does not change the dynamics of the theory, at least at the classical level. We study the spontaneous symmetry breaking for this new Lagrangian and show that how it works in Abelian and non-Abelian gauge theories. In the case of Abelian gauge theory our method adds nothing new to the so-called Higgs mechanism. However, in the non-Abelian case topological degrees of freedom, as classical fields, arise. Finally, we reacquire our results considering a new definition for the vacuum.
\end{abstract}

\section{Introduction} \label{sec1}

Spontaneous symmetry breaking is at the heart of the Standard Model of particle physics. It is an important component in understanding the origin of elementary particle masses. According to Goldstone theorem \cite{Goldstone:1962es}, if a continuous global symmetry is broken spontaneously, for each broken group generator there must appear in the theory a massless particle called Nambu-Goldstone boson. However, in the case of local symmetries, one can evade Goldstone theorem using Higgs mechanism \cite{Higgs:1964pj,Englert:1964et,Guralnik:1964eu}. In this mechanism some gauge bosons get mass and a massive spinless particle, Higgs particle, appears in the theory.

Shortly after discovering the Higgs mechanism in the 1960s, a new approach to quantum field theory developed and became common in the 1970s. Some physicists began to interpret some of the solutions of the classical field equations as candidates for particles of the theory. This particles are different from the elementary particles that arise from the quantization of the fields. The main difference is the topological structure of this new, particle-like solutions which differ from the vacuum. Interestingly, these solutions of the classical field equations, topological solitons, appear when spontaneous symmetry breaking occurs in the quantum field theory level, for example, see the Nielsen-Olesen vortex solution in Abelian-Higgs model \cite{Nielsen:1973cs} and 't Hooft-Polyakov monopole solution in SU(2) Georgi-Glashow model \cite{tHooft:1974kcl,Polyakov:1974ek}. In both cases, spontaneous symmetry breaking can occur and, on the other hand, topological solutions exist. This induce the possibility of a new mechanism for symmetry breaking in which, both particle and topological degrees of freedom can appear. The theories which describe both topological and particle degrees of freedom is not new. Indeed, there is a formulation of the pure Yang-Mills theory in terms of new variables reflecting the topological degrees of freedom. This formulation is known as Cho decomposition in which the Yang-Mills field is decomposed into other fields \cite{Cho:1979nv,Cho:1980nx}.

Cho decomposition, along with Abelian projection \cite{tHooft:1981bkw}, is a way to extract topological degrees of freedom in the pure Yang-Mills theory. Unlike Abelian projection, which is a partial gauge fixing method, in Cho decomposition topological defects emerge without gauge fixing \cite{Shabanov:1999xy,Shabanov:1999uv}. It is supposed that topological degrees of freedom dominate the low-energy limit of Yang-Mills theories. Indeed, there are many models in which the vacuum of the Yang-Mills theory is filled with topological solitons such as vortices and monopoles. These topological objects give structure to the vacuum and they can describe low energy phenomena like color confinement which can not understood with perturbative methods that is quantum particles and their interactions. In Cho's restricted decomposition for SU(2) Yang-Mills field, there are four degrees of freedom: two dynamical and two topological. Cho's decomposition has been developed by Faddeev and Niemi \cite{Faddeev:1998eq}. In Faddeev-Niemi decomposition, knotlike solitons can appear in low-energy limit of SU(2) pure Yang-Mills theory.

In this paper, we introduce a procedure for decomposing both scalar field and gauge field in U(1) and SU(2) Yang-Mills-Higgs theory. According to this decomposition, we rewrite the Lagrangian based on new variables which does not change the Euler-Lagrange equations. Considering these new variables and using some constraints on the classical fields, vacuum constraints, one can reobtain Cho's restricted and extended theory from SU(2) Yang-Mills-Higgs Lagrangian \cite{Mohamadnejad:2015vua}. In our approach the topological field $ \textbf{n} $ which is the orientation of the scalar field in color space, in addition to the particle degrees of freedom, appear in the Lagrangian. In Cho decomposition of SU(2) Yang-Mills field, the extra degrees of freedom induced by $ \textbf{n} $ were puzzling. It was tried to demolish them by extra constraints \cite{Shabanov:1999xy,Shabanov:1999uv}. On the other hand, some authors interpret the field $ \textbf{n} $ as a dynamical field \cite{Faddeev:1998eq}. However, these interpretations has been criticized in \cite{Bae:2001uw}. We consider this field as a topological field which is present in the vacuum and makes it nontrivial and other degrees of freedom are quantum fields in this vacuum which now has structure duo to the topological field $ \textbf{n} $. It is remarkable that in reformulated Lagrangian, the topological field $ \textbf{n} $ is present even after symmetry breaking. Albeit, topological field only appear in non-Abelian theory. We show that topological field disappears in Abelian-Higgs model. Therefore, our symmetry breaking approach leads to the same result as Higgs mechanism in the case of Abelian theory. But, in non-Abelain case, our approach is different from Higgs. The vacuum in Higgs approach is empty form topological fields, while vacuum in our method, for non-Abelian case, is filled by topological field as a classical background. Hence, the vacuum of a non-Abelain gauge theory is much peculiar than Abelain one.

Eventually, we reacquire our results for symmetry breaking in reformulated Yang-Mills-Higgs theory considering a new definition for the vacuum. In Higgs mechanism, a constant universal field as a vacuum expectation value, is present in whole space or vacuum. In addition to this (constant) vacuum field, we also allow that gauge fields without matter source be present in the vacuum as classical (vacuum) fields. Our motivation is this classical assumption that classical fields like electromagnetic and gravitational field can be extended in whole space, and so vacuum is not necessarily empty of classical fields. It should be only empty of matter fields or particles. In other words, classical fields as topological degrees of freedom can be present and form the vacuum or space. We also suppose that, in addition to the potential term of the Higgs sector, the kinetic term is minimum in the vacuum, too. According to this assumption (vacuum) gauge field automatically would be without matter source in Yang-Mills-Higgs theories. Quantum particles are excitations above these scalar and gauge (vacuum) fields. Considering this revision of the vacuum, spontaneous symmetry breaking leads to the same result of the reformulated Yang-Mills-Higgs theory and provides another interpretation for our approach.

In the next section, Sec. \ref{sec2}, we introduce new variables for U(1) and SU(2) Yang-Mills-Higgs theory. We write Lagrangian based on these new variables and show that Euler-Lagrange equations do not change for these variables and therefore the dynamics of the theory remains the same, at least at the classical level. In Sec. \ref{sec3} we study spontaneous symmetry breaking for reformulated theory, and for the non-Abelian case, we show that after symmetry breaking extra degrees of freedom appear in the theory. These extra degrees of freedom are nothing but the topological ones and they should not be interpreted as quantum fields associated with particles. They are simply classical background fields which give structure to the vacuum and make it non-trivial. In Sec. \ref{sec4}, a new interpretation for our results is presented and we reacquire the same results by redefinition of the vacuum. Finally, the conclusion is given in Sec. \ref{sec5}.

\section{Yang-Mills-Higgs theory in new variables} \label{sec2}

We first consider the Abelian-Higgs model with the following Lagrangian
\begin{equation}
L = \frac{1}{2} (D_{\mu} \phi)^{*} (D^{\mu} \phi) - V(\phi^{*} \phi)  - \frac{1}{4} F_{\mu\nu} F^{\mu\nu} .  \label{eq001}
\end{equation}
where
\begin{eqnarray}
D_{\mu} \phi  &=& \partial_{\mu} \phi  + ig  A_{\mu} \phi   , \nonumber\\
F_{\mu\nu} &=& \partial_{\mu} A_{\nu} - \partial_{\nu} A_{\mu}   , \nonumber\\
V(\phi^{*} \phi) &=&  \frac{\lambda}{4} (\phi^{*} \phi  -  \nu^{2})^{2} , \quad \quad \lambda \, , \, \nu > 0  . \label{eq002}
\end{eqnarray}
Euler-Lagrange equations for this model are:
\begin{eqnarray}
 \partial_{\nu} F^{\mu\nu} &=& - \frac{ig}{2} (\phi^{*} (D^{\mu} \phi) - \phi (D^{\mu} \phi)^{*} )  , \label{eq003}  \\
D_{\mu} D^{\mu} \phi &=&  - \lambda {\phi}  (\phi^{*} \phi  -  \nu^{2}) . \label{eq004}
\end{eqnarray}
The scalar field is a complex field with two components where in polar coordinate can be written as
\begin{equation}
\phi(x) = \rho(x) e^{i\theta(x)} .  \label{eq005}
\end{equation}
Substituting $ \phi = \rho e^{i\theta} $ in covariant derivative $ D_{\mu} \phi $, we get
\begin{equation}
D_{\mu} \phi  = e^{i\theta} \partial_{\mu} \rho + \rho D_{\mu} e^{i\theta}  ,  \label{eq006}
\end{equation}
where
\begin{eqnarray}
&& D_{\mu} e^{i\theta} = i (\partial_{\mu} \theta + g A_{\mu}) e^{i\theta} ,  \nonumber\\
&\Rightarrow& - i e^{-i\theta} D_{\mu} e^{i\theta} = \partial_{\mu} \theta + g A_{\mu} ,  \nonumber\\
&\Rightarrow& A_{\mu} = - \frac{1}{g} \partial_{\mu} \theta - \frac{i}{g} e^{-i\theta} D_{\mu} e^{i\theta} \label{eq007}
\end{eqnarray}
Introducing the field $ C_{\mu} $ so that
\begin{equation}
C_{\mu}  =  - \frac{i}{g} e^{-i\theta} D_{\mu} e^{i\theta}  ,  \label{eq008}
\end{equation}
we get
\begin{equation}
A_{\mu} = - \frac{1}{g} \partial_{\mu} \theta +  C_{\mu} .  \label{eq009}
\end{equation}
Note that eq. (\ref{eq009}) is the same as U(1) gauge transformation.

Now we change the variables of the model from old ones $ \phi $ and $ A_{\mu} $ to new ones $ \rho $, $ \theta $, and $ C_{\mu} $ where
\begin{equation}
\phi = \rho e^{i\theta} , \quad A_{\mu} = - \frac{1}{g} \partial_{\mu} \theta +  C_{\mu} . \label{eq010}
\end{equation}
In terms of new variables we have
\begin{eqnarray}
D^{\mu} \phi &=& (\partial^{\mu} \rho + ig C^{\mu}  \rho) e^{i\theta} ,   \nonumber\\
D_{\mu} D^{\mu} \phi &=&  (\partial_{\mu}  \partial^{\mu} \rho - g^{2} C_{\mu} C^{\mu}  \rho + i g [\rho \partial_{\mu} C^{\mu} + 2 C^{\mu}  \partial_{\mu} \rho ]) e^{i\theta} , \nonumber\\
F_{\mu\nu} &=& \partial_{\mu} C_{\nu} - \partial_{\nu} C_{\mu}   ,  \label{eq011}
\end{eqnarray}
and the Euler-Lagrange equations will be
\begin{eqnarray}
\partial_{\nu} F^{\mu\nu} &=& g^{2} \rho^{2} C^{\mu} , \label{eq012}  \\
\partial_{\mu}  \partial^{\mu} \rho - g^{2} C_{\mu} C^{\mu}  \rho  &=&  - \lambda \rho (\rho^{2}  -  \nu^{2}) , \label{eq013} \\
\rho \partial_{\mu} C^{\mu} + 2 C^{\mu}  \partial_{\mu} \rho &=& 0 . \label{eq014}
\end{eqnarray}
Note that Eq. (\ref{eq014}) is not independent from  Eq. (\ref{eq012}):
\begin{equation}
\partial_{\mu} \partial_{\nu} F^{\mu\nu} = g^{2} \rho (\rho \partial_{\mu} C^{\mu} + 2 C^{\mu}  \partial_{\mu} \rho) = 0 . \label{eq015}
\end{equation}
In terms of new variables Lagrangian (\ref{eq001}) will be
\begin{equation}
L = \frac{1}{2} \partial_{\mu} \rho \partial^{\mu} \rho + \frac{1}{2} g^{2} \rho^{2} C_{\mu} C^{\mu} - \frac{\lambda}{4} (\rho^{2}  -  \nu^{2})^{2}  - \frac{1}{4} F_{\mu\nu} F^{\mu\nu} . \label{eq016}
\end{equation}
Eqs. (\ref{eq012}) and (\ref{eq013}) can be obtained from the Lagrangian (\ref{eq016}).
Therefore, our reformulation does not change the Euler-Lagrange equations.
Interestingly, Lagrangian (\ref{eq016}), as well as Euler-Lagrange equations (\ref{eq012}) and (\ref{eq013}), does not contain the real field $ \theta(x) $; it only contains $ \rho(x) $ and $ C_{\mu}(x) $. Hence, $ \theta(x) $ is not a dynamical field and it does not contribute to energy-momentum of the model.

Now we reformulate SU(2) Yang-Mills-Higgs model with the Lagrangian:
\begin{equation}
L = \frac{1}{2} D_{\mu} \bm{\phi} \, . \, D^{\mu} \bm{\phi}  - V(\bm{\phi} \, . \, \bm{\phi})  -\frac{1}{4} \textbf{F}_{\mu\nu} \,. \, \textbf{F}^{\mu\nu}  , \label{eq017}
\end{equation}
where
\begin{eqnarray}
D_{\mu} \bm{\phi}  &=& \partial_{\mu} \bm{\phi}  + g  \textbf{A}_{\mu} \times \bm{\phi}   , \nonumber\\
\textbf{F}_{\mu\nu} &=& \partial_{\mu} \textbf{A}_{\nu} - \partial_{\nu} \textbf{A}_{\mu} + g \textbf{A}_{\mu} \times \textbf{A}_{\nu}  , \nonumber\\
V(\bm{\phi} \, . \, \bm{\phi})  &=&  \frac{\lambda}{4} (\bm{\phi} \, . \, \bm{\phi}  -  \nu^{2})^{2} , \quad \quad \lambda \, , \, \nu > 0  , \label{eq018}
\end{eqnarray}
and the Euler-Lagrange equations are:
\begin{eqnarray}
D_{\nu} \textbf{F}^{\mu\nu} &=& g \bm{\phi}  \times   D^{\mu} \bm{\phi}   , \label{eq019} \\
D_{\mu} D^{\mu} \bm{\phi}  &=&  - \lambda \bm{\phi}  (\bm{\phi} \, . \, \bm{\phi}  -  \nu^{2}) . \label{eq020}
\end{eqnarray}

Since the scalar field $ \bm{\phi} $ is a vector in 3D internal space, therefore, it has a magnitude and a direction and can be written as
\begin{equation}
\bm{\phi} =   \phi \, \textbf{n}  \, , \quad  (\textbf{n} \, . \, \textbf{n} = 1) . \label{eq021}
\end{equation}
Note that $ \phi $ has the magnitude and dimension of $ \bm{\phi} $, and $ \textbf{n} $ is a dimensionless unit field having the direction of $ \bm{\phi} $.
Covariant derivative for $ \bm{\phi} =   \phi \, \textbf{n} $ will be
\begin{equation}
D_{\mu} \bm{\phi} =  (\partial_{\mu} \phi) \textbf{n} + \phi D_{\mu} \textbf{n}     , \label{eq022}
\end{equation}
where
\begin{eqnarray}
D_{\mu} \textbf{n} &=&  \partial_{\mu} \textbf{n} + g \textbf{A}_{\mu} \times \textbf{n}  , \nonumber  \\
&\Rightarrow& \textbf{n} \times D_{\mu} \textbf{n} = \textbf{n} \times  \partial_{\mu} \textbf{n}
+g \textbf{A}_{\mu} - g (\textbf{A}_{\mu} . \textbf{n}) \textbf{n} , \nonumber  \\
&\Rightarrow& \textbf{A}_{\mu} =  (\textbf{A}_{\mu} . \textbf{n}) \textbf{n} + \frac{1}{g} \partial_{\mu} \textbf{n} \times \textbf{n} +
\frac{1}{g} \textbf{n} \times D_{\mu} \textbf{n}  . \label{eq023}
\end{eqnarray}
Introducing two new variables, $ A_{\mu} $ and $ \textbf{X}_{\mu} $, so that
\begin{eqnarray}
A_{\mu} &=&  \textbf{A}_{\mu} . \textbf{n}  , \nonumber  \\
\textbf{X}_{\mu} &=&  \frac{1}{g} \textbf{n} \times D_{\mu} \textbf{n} \, , \quad  (\textbf{X}_{\mu} \, . \, \textbf{n} = 0) , \label{eq024}
\end{eqnarray}
we get
\begin{equation}
\textbf{A}_{\mu} = A_{\mu}  \textbf{n} + \frac{1}{g} \partial_{\mu} \textbf{n} \times \textbf{n} +  \textbf{X}_{\mu} . \label{eq025}
\end{equation}
The Eq. (\ref{eq025}) is the same as Cho extended decomposition for SU(2) Yang-Mills field \cite{Cho:1980nx}.

Changing the variables of the model from the original ones $ \bm{\phi} $ and $ \textbf{A}_{\mu} $ to the new ones $ \phi $, $ \textbf{n} $, $ A_{\mu} $ and $ \textbf{X}_{\mu} $ where
\begin{eqnarray}
\bm{\phi} &=&  \phi \textbf{n}  , \nonumber  \\
\textbf{A}_{\mu} &=& A_{\mu}  \textbf{n} + \frac{1}{g} \partial_{\mu} \textbf{n} \times \textbf{n} +  \textbf{X}_{\mu} , \label{eq026}
\end{eqnarray}
with these constraints
\begin{equation}
\textbf{n} . \textbf{n} = 1 \, , \quad \textbf{X}_{\mu} \, . \, \textbf{n} = 0 , \label{eq027}
\end{equation}
we get
\begin{eqnarray}
&& D^{\mu} \bm{\phi} =  (\partial^{\mu} \phi) \textbf{n} + g \phi \textbf{X}^{\mu} \times \textbf{n}   , \nonumber  \\
&& D_{\mu} D^{\mu} \bm{\phi} =   (\partial_{\mu} \partial^{\mu} \phi - g^{2} \phi \textbf{X}_{\mu} . \textbf{X}^{\mu}  ) \textbf{n}
+\frac{g}{\phi} [\partial_{\mu} (\phi^{2} \textbf{X}^{\mu}) + g \phi^{2} A_{\mu} \textbf{n} \times \textbf{X}^{\mu} ] \times \textbf{n} , \nonumber  \\
&& \textbf{F}_{\mu\nu} = \widehat{\textbf{F}}_{\mu\nu} +  \widehat{D}_{\mu} \textbf{X}_{\nu} -  \widehat{D}_{\nu} \textbf{X}_{\mu}
+ g \textbf{X}_{\mu} \times \textbf{X}_{\nu} , \label{eq028}
\end{eqnarray}
where
\begin{eqnarray}
\widehat{\textbf{F}}_{\mu\nu} = [ (\partial_{\mu} A_{\nu} - \partial_{\nu} A_{\mu}) - \frac{1}{g} \textbf{n} . (\partial_{\mu} \textbf{n} \times \partial_{\nu} \textbf{n} ) ] \textbf{n}   , \nonumber  \\
\widehat{D}_{\mu} \textbf{X}_{\nu} =  \partial_{\mu} \textbf{X}_{\nu} + g (A_{\mu} \textbf{n} + \frac{1}{g} \partial_{\mu} \textbf{n} \times \textbf{n}) \times \textbf{X}_{\nu} . \label{eq029}
\end{eqnarray}
Euler-Lagrange equations with respect to new variables are
\begin{eqnarray}
&& D_{\nu} \textbf{F}^{\mu\nu} = g^{2}  \phi^{2} \, \textbf{X}^{\mu}   , \label{eq030} \\
&& \partial_{\mu} \partial^{\mu} \phi - g^{2} \phi \, \textbf{X}_{\mu} \, . \, \textbf{X}^{\mu} =  - \lambda \phi  (\phi^{2}  -  \nu^{2}) , \label{eq031} \\
&& D_{\mu} [\phi^{2} \textbf{X}^{\mu}] = 0  . \label{eq032}
\end{eqnarray}
Note that Eq. (\ref{eq032}) can be derived from Eq. (\ref{eq030}):
\begin{equation}
D_{\nu} \textbf{F}^{\mu\nu} = g^{2}  \phi^{2} \, \textbf{X}^{\mu} \quad \Rightarrow \quad
D_{\mu} D_{\nu} \textbf{F}^{\mu\nu} = g^{2} D_{\mu} \,  [\phi^{2} \textbf{X}^{\mu}] = 0  , \label{eq033}
\end{equation}
but Eq. (\ref{eq030}) and Eq. (\ref{eq031}) are independent equations.
Lagrangian (\ref{eq017}) based on new variables is:
\begin{eqnarray}
L &=& \frac{1}{2}  (\partial_{\mu} \phi) (\partial^{\mu} \phi) + \frac{1}{2} g^{2} \phi^{2} \textbf{X}_{\mu} \, . \, \textbf{X}^{\mu} , \nonumber \\
& & - \frac{1}{4} \textbf{F}_{\mu\nu} \, . \, \textbf{F}^{\mu\nu} - \frac{\lambda}{4} (\phi^{2}  -  \nu^{2})^{2} . \label{eq034}
\end{eqnarray}
Variation with respect to the new variables $ A_{\mu} $, $ \textbf{X}_{\mu} $, and $ \phi $ leads to the following Euler-Lagrange equations, respectively:
\begin{eqnarray}
\textbf{n} \, . \, D_{\nu} \textbf{F}^{\mu\nu} &=& 0 , \label{eq035} \\
D_{\nu} \textbf{F}^{\mu\nu} &=& g^{2}  \phi^{2} \, \textbf{X}^{\mu} , \label{eq036}  \\
\partial_{\mu} \partial^{\mu} \phi - g^{2} \phi \, \textbf{X}_{\mu} \, . \, \textbf{X}^{\mu} &=& - \lambda \phi  (\phi^{2}  -  \nu^{2}) ,  \label{eq037}
\end{eqnarray}
and variation with respect to $ \textbf{n} $ yields a trivial identity.
Regarding $ \textbf{X}_{\mu} \, . \, \textbf{n} = 0  $, one can also see that Eq. (\ref{eq035}) can be derived from Eq. (\ref{eq036}).
Hence, we left with Eqs. (\ref{eq036}) and (\ref{eq037}) which are the same as Eqs. (\ref{eq030}) and (\ref{eq031}). This means that our reformulation has not changed the dynamics of the model, at least at the classical level.

In this section we started with two models with Lagrangians (\ref{eq001}) and (\ref{eq017}) and reformulated them with new variables. We also show that our reformulations lead to the same Euler-Lagrange equations. Therefore, the dynamics of our reformulations is the same as the original models.
In the next section, we study spontaneous symmetry breaking for reformulated Lagrangians, Eqs. (\ref{eq016}) and (\ref{eq034}), and show that for the Abelian case, it leads to the same result as the Higgs mechanism, while for the non-Abelain case, extra degrees of freedom arise.

\section{Symmetry breaking in reformulated Yang-Mills-Higgs theory} \label{sec3}

Consider again the Lagrangian (\ref{eq001}) for the Abelian-Higgs model. For this case we have continuous degenerate vacuum state at $ \phi^{*}\phi = \nu^{2} $.
By choosing one of these degenerate vacua,  for example $ \nu $, and setting $ \phi = \nu + \phi_{1} + i \phi_{2} $, we get a Lagrangian in terms of the new fields
 $ \phi_{1} $ and $ \phi_{2} $. Doing this, the gauge field $ A_{\mu} $ becomes massive. The scalar field $ \phi_{1} $ also gets mass, but $ \phi_{2} $ appears to be a massless field. However, the Nambu-Goldstone boson $ \phi_{2} $, can be eliminated by a gauge transformation . In this gauge, unitary gauge, the Lagrangian contains only two massive physical fields, the gauge field with spin 1, and $ \phi_{1} $ with spin 0. This is the Higgs mechanism in which
the $ \phi_{2} $ field that in the case of spontaneous symmetry breaking of the global symmetry became massless has disappeared, and in addition, the gauge field has now acquired a mass.

Now we study spontaneous symmetry breaking for the reformulation of the Lagrangian (\ref{eq001}). Consider our reformulated Lagrangian (\ref{eq016}):
\begin{equation}
L = \frac{1}{2} \partial_{\mu} \rho \partial^{\mu} \rho + \frac{1}{2} g^{2} \rho^{2} C_{\mu} C^{\mu} - \frac{\lambda}{4} (\rho^{2}  -  \nu^{2})^{2}  - \frac{1}{4} F_{\mu\nu} F^{\mu\nu} . \nonumber
\end{equation}
In this Lagrangian there is no continuous vacuum state. There are just two discrete vacuum states $ \nu $ and $ -\nu $. By choosing  $ \nu $ as the vacuum expectation value and putting $ \rho \rightarrow \nu + \rho  $ we get
\begin{eqnarray}
L &=& - \frac{1}{4} F_{\mu\nu} F^{\mu\nu}
+ \frac{1}{2} m^{2}_{C} C_{\mu} C^{\mu}   \nonumber\\
 &+& \nu g^{2} \rho C_{\mu} C^{\mu} + \frac{g^{2}}{2} \rho^{2}  C_{\mu} C^{\mu}  \nonumber\\
 &+& \frac{1}{2} \partial_{\mu} \rho \, \partial^{\mu} \rho - \frac{1}{2} m^{2}_{\rho} \rho^{2} - \lambda \nu \rho^{3} - \frac{\lambda}{4}  \rho^{4}   ,  \label{eq038}
\end{eqnarray}
where $ m_{\rho} = \sqrt{2 \lambda \nu^{2}} $ and $ m_{C} = g \nu $. This is the same result as the Higgs mechanism. The difference is that we did not choose a special gauge, unitary gauge, in order to eliminate Nambu-Goldstone boson. Indeed in Lagrangian (\ref{eq016}) there is only one scalar field which after spontaneous symmetry breaking becomes massive.

Spontaneous symmetry breaking for our reformulation of the Abelian-Higgs model yielded the same result as the Higgs mechanism. However, in the case of SU(2) Yang-Mills-Higgs model, Spontaneous symmetry breaking for our reformulated Lagrangian (\ref{eq034}) leads to new result. In this case apart from usual particle degrees of freedom which also appear in Higgs mechanism, topological degrees of freedom as classical fields arise.
In Lagrangian (\ref{eq034}):
\begin{eqnarray}
L &=& \frac{1}{2}  (\partial_{\mu} \phi) (\partial^{\mu} \phi) + \frac{1}{2} g^{2} \phi^{2} \textbf{X}_{\mu} \, . \, \textbf{X}^{\mu} , \nonumber \\
& & - \frac{1}{4} \textbf{F}_{\mu\nu} \, . \, \textbf{F}^{\mu\nu} - \frac{\lambda}{4} (\phi^{2}  -  \nu^{2})^{2} , \nonumber
\end{eqnarray}
there are two discrete degenerate vacuum states.
We choose $ \nu $ as the vacuum expectation value and by setting $ \phi \rightarrow \nu + \phi $ Lagrangian becomes
\begin{eqnarray}
L &=& - \frac{1}{4} \textbf{F}_{\mu\nu} . \textbf{F}^{\mu\nu}
+ \frac{1}{2} m^{2}_{X} \textbf{X}_{\mu} \, . \, \textbf{X}^{\mu}  \nonumber\\
 &+& \nu g^{2} \phi \textbf{X}_{\mu} \, . \, \textbf{X}^{\mu} + \frac{g^{2}}{2} \phi^{2}  \textbf{X}_{\mu} \, . \, \textbf{X}^{\mu}  \nonumber\\
 &+& \frac{1}{2} \partial_{\mu} \phi \, \partial^{\mu} \phi - \frac{1}{2} m^{2}_{\phi} \phi^{2} - \lambda \nu \phi^{3} - \frac{\lambda}{4}  \phi^{4}   ,  \label{eq039}
\end{eqnarray}
where $ m_{\phi} = \sqrt{2 \lambda \nu^{2}} $ and $ m_{X} = g \nu $.
Keep in mind that Lagrangian (\ref{eq034}) as well as Lagrangian (\ref{eq039}) is invariant under following infinitesimal transformations:
\begin{eqnarray}
&& \delta \textbf{n} = - \textbf{a} \times \textbf{n}, \nonumber \\
&& \delta \textbf{X}_{\mu} = - \textbf{a} \times \textbf{X}_{\mu},  \nonumber \\
&& \delta A_{\mu} = \frac{1}{g} \textbf{n} . \partial _{\mu} \textbf{a}.  \label{eq040}
\end{eqnarray}
In Lagrangian (\ref{eq039}), apart from particle degrees of freedom, which are one massive scalar $ \phi $, one massless vector $ A_{\mu} $, and two massive vectors $ \textbf{X}_{\mu} $, topological degrees of freedom associated with $ \textbf{n} $ emerges.
If $ \textbf{n} $ was not a function of space-time, then our result would become the same as the Higgs mechanism.
It is also interesting that the potential part of the Lagrangian  (\ref{eq034}):
\begin{eqnarray}
V(\phi,A_{\mu},\textbf{X}_{\mu}) &=& \frac{\lambda}{4} (\phi^{2} - \nu^{2})^{2}
 - \frac{1}{2} g^{2} \phi^{2} \textbf{X}_{\mu} \, . \, \textbf{X}^{\mu} \nonumber\\
 &+&  \frac{1}{2} g^{2} (A_{\mu} A^{\mu}  \textbf{X}_{\nu} \, . \, \textbf{X}^{\nu} - A_{\mu} A^{\nu}  \textbf{X}_{\nu} \, . \, \textbf{X}^{\mu} )  \nonumber\\
 &+& \frac{1}{4} g^{2} (\textbf{X}_{\mu} \times \textbf{X}_{\nu}) .  (\textbf{X}^{\mu} \times \textbf{X}^{\nu}) ,  \label{eq041}
\end{eqnarray}
will be minimum for $ \phi = \nu $ and $ \textbf{X}_{\mu} = 0 $; and since there is no constraint on $ A_{\mu} $ it could get any value.
Note that $ A_{\mu} $ behaves differently from $ \phi $ and $ \textbf{X}_{\mu} $. These fields, $ \phi $ and $ \textbf{X}_{\mu} $, can take vacuum expectation value, $ \langle \phi \rangle = \nu $ and $ \langle \textbf{X}_{\mu} \rangle = 0 $, while there is no vacuum constraint on $ A_{\mu} $. In addition, according to (\ref{eq040}), $ \bm{\phi} $ and $ \textbf{X}_{\mu} $ transform the same under local rotation of internal space, and again $ A_{\mu} $ transforms in a different manner. Another difference between $ \phi $, $ \textbf{X}_{\mu} $ and $ A_{\mu} $ is that $ A_{\mu} $ unlike $ \phi $ and $ \textbf{X}_{\mu} $ remains massless after spontaneous symmetry breaking.
The Lagrangian  (\ref{eq034}) for the vacuum states,  $ \langle \phi \rangle = \nu $ and $ \langle \textbf{X}_{\mu} \rangle = 0 $, is
\begin{eqnarray}
 L_{V} &=& - \frac{1}{4} \widehat{\textbf{F}}_{\mu\nu}^{2}  \nonumber\\
 &=& - \frac{1}{4} [ (\partial_{\mu} A_{\nu} - \partial_{\nu} A_{\mu}) - \frac{1}{g} \textbf{n} . (\partial_{\mu} \textbf{n} \times \partial_{\nu} \textbf{n} ) ]^{2} .  \label{eq042}
\end{eqnarray}
We call $ L_{V} $ "vacuum Lagrangian" and we assume that massless field $ A_{\mu} $ as well as topological field $ \textbf{n} $ can be present in the vacuum.
Note that Lagrangian (\ref{eq042}) is the same as the Lagrangian of Cho's restricted gauge theory \cite{Cho:1979nv} which is proposed to describe low energy properties associated with vacuum structure of non-Abelian gauge theories.

Although the topological field $ \textbf{n} $ is not a dynamical field -variation of the Lagrangian (\ref{eq034}) with respect to $ \textbf{n} $ leads to a trivial identity- however, it carries energy and momentum and contributes in vacuum energy.
Notice that for the non-Abelain case, we have relaxed the traditional condition $ \textbf{F}_{\mu \nu} = 0 $ for the vacuum. 
Observationally, the possibility of $ \textbf{F}_{\mu \nu} \neq 0 $ is not strange – because of the cosmic microwave background
radiation.
There are many motivations for the non-zero energy of the vacuum. For example, some models for color confinement suppose that vacuum is filled with objects having energy such as vortices, monopoles and knot-like solitons. Another motivation is the late time accelerating expansion of the universe associated with dark energy.
Efforts to interpret dark energy as quantum fluctuations, zero point energy, have been drastically failed and therefore new approaches are welcome.
One of these new approaches is introducing a field, scalar or even vector field, which is coupled to gravity and can explain accelerating expansion of the universe.
On the other hand, in classical physics, all objects are either (force) field or matter, and the big difference between them is that classical (force) fields, electromagnetic and gravitational fields, unlike matter, which is localized, can be everywhere outside of their matter sources. Indeed, we can define classical vacuum as a
space which is not empty in the sense that (force) fields, unlike their matter sources, can be present there. Hence, vacuum (force) fields are the solutions of field equations without matter source.
In particle physics, gauge fields play the role of force fields. Therefore, we propose that they are allowed in vacuum without matter sources.
In the next section we examine this proposal and we show that it leads to the same result for the symmetry breaking.

\section{New approach based on redefinition of the vacuum} \label{sec4}

In gauge field theories, a vacuum solution is a solution of a field equation in which the sources of the field are taken to be identically zero.
For example, in Maxwell's theory of electromagnetism, a vacuum solution would represent the electromagnetic field in a region of space where there are no electromagnetic sources:
\begin{equation}
\partial_{\mu} F^{\mu\nu} = J^{\nu} = 0 .  \label{eq01}
\end{equation}

On the other hand, in (complex) scalar field theory the Lagrangian made of kinetic and potential terms:
\begin{equation}
L = \frac{1}{2} \partial_{\mu} \phi^{*} \partial^{\mu} \phi - V(\phi^{*} \phi) ,  \label{eq02}
\end{equation}
and the vacuum state can be obtained by minimizing the potential term $ V(\phi^{*} \phi) $.
If the minimum occurs at $ \phi^{*} \phi = \nu^{2} $, One should choose a vacuum state from the set of degenerate vacua $ \phi^{V} = \nu e^{i \theta_{V}} $ related to each other by rotation. This vacuum has chosen for all space-time points and it is constant:
\begin{equation}
\phi^{V}  = constant .  \label{eq03}
\end{equation}
In addition, the physical fields $ \phi^{phys}(x) $ are excitations above the vacuum
\begin{equation}
\phi(x) \rightarrow  \phi^{V} + \phi^{phys}(x).  \label{eq04}
\end{equation}
Note that $ \phi^{V} $ is a (constant) classical background field and $ \phi^{phys}(x) $ is a quantum field associated with particles. The Lagrangian based on the new field $ \phi^{phys}(x) $ is not invariant under the same transformation of the old field $ \phi(x) \rightarrow \phi(x) e^{i \theta}  $. This phenomenon is known as spontaneous symmetry breaking.
According to the Goldstone theorem massless particles, known as Nambu-Goldstone bosons, are unavoidable in spontaneously symmetry broken scalar field theories \cite{Goldstone:1962es}.
However, Higgs and others pointed out that in gauge theories it is possible to evade Goldstone theorem \cite{Higgs:1964pj,Englert:1964et,Guralnik:1964eu}.
In these theories, according to Higgs mechanism, gauge fields can get mass and there is no room for Nambu-Goldstone boson.

Until now, there were two conditions for vacuum. One came from gauge field theory Eq. (\ref{eq01}), and the other was minimizing the potential term in scalar field theory. We consider both of them as a definition of vacuum in gauge theory including both scalar and gauge fields. We add another condition according to which one has to minimize the scalar part of the Hamiltonian density derived from symmetric and gauge invariant energy-momentum tensor ($ H = T_{00} $). This means that not only the potential, but also the kinetic term of the scalar field should be minimized. Furthermore, vacuum states clearly should be a solution of Euler-Lagrange equation and they can be a field in general, so we treat them as the classical (background) fields and we relax the condition (\ref{eq03}). Therefore, every space-time point can have its own vacuum state, at least on cosmological scales.

Synoptically, we follow these steps in order to break the symmetry in gauge theories including both scalar and gauge fields:

1- Scalar part of the Hamiltonian density $ H^{S} = H(D_{\mu} \bm{\phi} , \bm{\phi}) $ should be minimized:
\begin{equation}
 \frac{\partial H^{S}}{\partial \bm{\phi}} = \frac{\partial H^{S}}{\partial (D_{\mu} \bm{\phi})} = 0 ,  \label{eq05}
\end{equation}
where the minimum exists if
\begin{eqnarray}
&&  \frac{\partial^{2} H^{S}}{\partial \bm{\phi}^{2}} > 0 \,   , \nonumber\\
&& [\frac{\partial^{2} H^{S}}{\partial \bm{\phi} \partial (D_{\mu} \bm{\phi})}]^{2} -
 \frac{\partial^{2} H^{S}}{\partial \bm{\phi}^{2}}  \frac{\partial^{2} H^{S}}{\partial (D_{\mu} \bm{\phi})^{2}} < 0 \, .  \label{eq06}
\end{eqnarray}

2- The source term in gauge field equation are taken to be zero:
\begin{equation}
D_{\mu} \textbf{F}^{\mu\nu} = \textbf{J}^{\nu} = 0 .  \label{eq07}
\end{equation}

3- Every space-time point has its own vacuum. In other words, we have vacuum fields. Moreover, the physical (quantum) fields, $ \bm{\phi}^{phys}(x) \, and \, \, \textbf{A}_{\mu}^{phys}(x) $, are excitations above the (classical) vacuum fields $ \bm{\phi}^{V}(x) \, and \, \, \textbf{A}_{\mu}^{V}(x) $:
\begin{eqnarray}
\bm{\phi}(x) &\rightarrow& \bm{\phi}^{V}(x) + \bm{\phi}^{phys}(x)   , \nonumber\\
\textbf{A}_{\mu}(x) &\rightarrow& \textbf{A}_{\mu}^{V}(x) + \textbf{A}_{\mu}^{phys}(x) .  \label{eq08}
\end{eqnarray}

We have supposed that gauge fields with non-zero energy in principle can be present in vacuum, so, in step 1, we only minimize the scalar part of the Hamiltonian density.
Now we are ready to see the consequences of our procedure for two examples: Abelian $ U(1) $ gauge theory and non-Abelian $ SU(2) $ gauge theory.

For the Abelian case with Lagrangian (\ref{eq001}), from the symmetric and gauge invariant energy-momentum tensor, the scalar part of the Hamiltonian density is
\begin{equation}
H^{S} = T^{S}_{00} = \frac{1}{2} (D_{0} \phi)^{*}(D_{0} \phi) + \frac{1}{2} (D_{i} \phi)^{*}(D_{i} \phi) + \frac{\lambda}{4}(\phi^{*} \phi - \nu^{2})^{2} .  \label{eq12}
\end{equation}
where $ i $ run over the three spatial coordinate labels.
Obviously the Hamiltonian density (\ref{eq12}) will be minimized ($ H^{S}_{min} = 0 $) at
\begin{equation}
D_{\mu} \phi = 0 \, \, , \, \, \phi^{*} \phi  =  \nu^{2}  ,  \label{eq13}
\end{equation}
and according to Eq. (\ref{eq003}) for $ D_{\mu} \phi = 0 $ we get $ \partial_{\nu} F^{\mu\nu} = 0 $.
Vacuum fields should satisfy Eqs. (\ref{eq13}). Note that condition $ D_{\mu} \phi = 0 $ is by itself extremely strong. This condition leads to
\begin{eqnarray}
&&\phi^{*} \phi = constant , \label{eq14}  \\
&&A_{\mu} =  \frac{i}{2g \phi^{*} \phi} (\phi^{*} \partial_{\mu} \phi - \phi \partial_{\mu} \phi^{*} ) \Rightarrow F_{\mu\nu} = 0 . \label{eq15}
\end{eqnarray}
It is remarkable that condition  $ D_{\mu} \phi = 0 $ alone leads to Eq. (\ref{eq14}). According to Eq. (\ref{eq14}) for non-trivial vacuum $ \phi \neq 0 $ we get spontaneous symmetry breaking even without a potential term in Lagrangian. According to $ F_{\mu\nu} = 0 $, vacuum energy density, which is the total Hamiltonian density including both scalar and gauge fields, is zero. So we conclude that vacuum of the Abelian gauge theory is structure-less and vacuum fields do not carry energy and momentum in this case.

Working in polar co-ordinates $ \phi(x) = \rho(x) e^{i\theta(x)} $,
vacuum fields are:
\begin{equation}
\rho^{V} = \nu \, \, and \, \, A^{V}_{\mu}(x) =  - \frac{1}{g} \partial_{\mu}\theta(x)  .  \label{eq16}
\end{equation}
By substituting
\begin{equation}
\rho(x) \rightarrow \nu +  \rho(x)  \, \, and \, \, A_{\mu}(x) \rightarrow - \frac{1}{g} \partial_{\mu}\theta(x) + A_{\mu}(x)   ,  \label{eq17}
\end{equation}
Lagrangian (\ref{eq001}) will be
\begin{eqnarray}
L &=& - \frac{1}{4} F_{\mu\nu} F^{\mu\nu} + \frac{1}{2} m^{2}_{A} A_{\mu} A^{\mu}    \nonumber\\
 &+& \frac{1}{2} \partial_{\mu} \rho \, \partial^{\mu} \rho - \frac{1}{2} m^{2}_{\rho} \rho^{2} + coupling \, \, terms  .  \label{eq18}
\end{eqnarray}
where $ m_{\rho} = \sqrt{2 \lambda \nu^{2}} $ and $ m_{A} = g \nu $.
Lagrangian  (\ref{eq18}) contains two physical (quantum) fields only, vector field $ A_{\mu}(x) $ with spin 1, and scalar field $ \rho(x) $ with spin 0, and they are both massive. Note that the vacuum field $ \theta(x) $ has disappeared in this case and our result is in agreement with the Higgs mechanism and our previous approach  (\ref{eq038}).

Now we consider the non-Abelian SU(2) gauge theory with the Lagrangian  (\ref{eq017}).
The scalar part of the Hamiltonian density is
\begin{equation}
H^{S} = T^{S}_{00} = \frac{1}{2} (D_{0} \bm{\phi})\, . \, (D_{0} \bm{\phi}) + \frac{1}{2} (D_{i} \bm{\phi})\, . \, (D_{i} \bm{\phi}) + \frac{\lambda}{4}(\bm{\phi} \, . \, \bm{\phi} - \nu^{2})^{2} .  \label{eq120}
\end{equation}
$ H^{S} $ will be minimized ($ H^{S}_{min} = 0 $) at
\begin{equation}
D_{\mu} \bm{\phi} = 0 \, \, , \, \, \bm{\phi} \, . \, \bm{\phi}  =  \nu^{2}  ,  \label{eq23}
\end{equation}
and according to Eq. (\ref{eq019}), the condition $ D_{\mu} \bm{\phi} = 0 $ implies $ D_{\nu} \textbf{F}^{\mu\nu} = 0 $.

Vacuum fields must fulfill Eqs. (\ref{eq23}). Regarding Eq. (\ref{eq021}), condition $ D_{\mu} \bm{\phi} = 0 $ leads to
\begin{eqnarray}
&& \bm{\phi} \, . \, \bm{\phi} = \phi^{2}  = constant , \label{eq25}  \\
&&\textbf{A}^{V}_{\mu} = A^{V}_{\mu}  \textbf{n} + \frac{1}{g} \partial_{\mu} \textbf{n} \times \textbf{n} , \label{eq26}
\end{eqnarray}
where $  A^{V}_{\mu} = \textbf{A}^{V}_{\mu} .\textbf{n}  $ is an unconstrained four-vector. We should mention that the vacuum form of gauge field $ \textbf{A}^{V}_{\mu} $ in Eq. (\ref{eq26}) is proposed before with different motivation \cite{Cho:1979nv}. Again according to the condition $ D_{\mu} \bm{\phi} = 0 $, for non-trivial vacuum $ \phi \neq 0 $, we can get spontaneous symmetry breaking without a potential term. Unlike the Abelian case, now we can have $ \textbf{F}_{\mu\nu}  \neq 0 $, so there exist vacuum energy density.
Therefore vacuum of the non-Abelian gauge theory has structure and vacuum fields carry energy and momentum.

Vacuum Lagrangian is
\begin{equation}
L_{V} = -\frac{1}{4} (F^{V}_{\mu\nu})^{2}  , \label{eq27}
\end{equation}
where
\begin{equation}
F^{V}_{\mu\nu} = [(\partial_{\mu} A^{V}_{\nu} - \partial_{\nu} A^{V}_{\mu}) - \frac{1}{g} \textbf{n} . (\partial_{\mu} \textbf{n} \times \partial_{\nu} \textbf{n} )]  . \label{eq28}
\end{equation}
By reparametrization of $ \textbf{n} $
\begin{equation}
\textbf{n} =
\begin{pmatrix}
\sin{\alpha} \, \cos{\beta} \\
\sin{\alpha} \, \sin{\beta} \\
\thickspace \cos{\alpha}
\end{pmatrix}
,  \label{eq29}
\end{equation}
where $ \alpha $ and $ {\beta} $ are fields, vacuum Lagrangian will be
\begin{equation}
L_{V} = -\frac{1}{4} (A_{\mu\nu} +B_{\mu\nu})  \, (A^{\mu\nu} + B^{\mu\nu})  , \label{eq30}
\end{equation}
where
\begin{eqnarray}
A_{\mu\nu} &=& \partial_{\mu} A^{V}_{\nu} - \partial_{\nu} A^{V}_{\mu} , \label{eq31}  \\
B_{\mu\nu} &=& - \frac{1}{g} \sin{\alpha} (\partial_{\mu} \alpha \, \partial_{\nu} \beta - \partial_{\nu} \alpha \, \partial_{\mu} \beta)  . \label{eq32}
\end{eqnarray}
Euler-Lagrange equations for vacuum fields $ A^{V}_{\mu} $, $ \alpha $, and $ \beta $ are
\begin{eqnarray}
\partial_{\mu}(A^{\mu\nu} + B^{\mu\nu}) &=& 0 , \label{eq33}  \\
\sin{\alpha} \, \partial_{\mu}[\partial_{\nu} \beta \, (A^{\mu\nu} + B^{\mu\nu} ) ] &=& 0 , \label{eq34}  \\
 \partial_{\mu} [ \sin{\alpha} \, \partial_{\nu} \alpha \, (A^{\mu\nu} + B^{\mu\nu} ) ] &=& 0 , \label{eq35}
\end{eqnarray}
respectively. Note that according to Eq. (\ref{eq33}), the other two equations, Eqs. (\ref{eq34}) and (\ref{eq35}), do not lead to new equations and there is only one equation for vacuum fields which is nothing but the vacuum condition (\ref{eq07}):
\begin{equation}
D_{\mu} \textbf{F}^{\mu\nu} = \partial_{\mu}(A^{\mu\nu} + B^{\mu\nu}) = 0 . \label{eq36}
\end{equation}

Now in order to obtain the complete Lagrangian we put
\begin{eqnarray}
&&\phi(x) \rightarrow \nu +  \phi(x) , \label{eq37} \\
&& \textbf{A}_{\mu}(x) \rightarrow A^{V}_{\mu}  \textbf{n} + \frac{1}{g} \partial_{\mu} \textbf{n} \times \textbf{n} + \textbf{A}^{phys}_{\mu}(x)  ,  \label{eq38}
\end{eqnarray}
where
\begin{equation}
\textbf{A}^{phys}_{\mu}(x)  =  A^{1}_{\mu} \textbf{n}_{1} + A^{2}_{\mu} \textbf{n}_{2} + A^{3}_{\mu} \textbf{n} , \label{eq39}
\end{equation}
and $ (\textbf{n}_{1},\textbf{n}_{2},\textbf{n}) $ forms an orthonormal basis for internal space.
By substituting (\ref{eq37}) and (\ref{eq38}) the Lagrangian (\ref{eq017}) will be
\begin{eqnarray}
L &=& - \frac{1}{4} \textbf{F}_{\mu\nu} . \textbf{F}^{\mu\nu}
+ \frac{1}{2} m^{2}_{A} [(A^{1}_{\mu})^{2} + (A^{2}_{\mu})^{2} ]    \nonumber\\
 &+& \nu g^{2} \phi  [(A^{1}_{\mu})^{2} + (A^{2}_{\mu})^{2} ] + \frac{g^{2}}{2} \phi^{2}  [(A^{1}_{\mu})^{2} + (A^{2}_{\mu})^{2} ]  \nonumber\\
 &+& \frac{1}{2} \partial_{\mu} \phi \, \partial^{\mu} \phi - \frac{1}{2} m^{2}_{\phi} \phi^{2} - \lambda \nu \phi^{3} - \frac{\lambda}{4}  \phi^{4}   ,  \label{eq40}
\end{eqnarray}
where $ m_{\phi} = \sqrt{2 \lambda \nu^{2}} $ and $ m_{A} = g \nu $ and
\begin{equation}
\textbf{F}_{\mu\nu} = \widehat{\textbf{F}}_{\mu\nu} +  \widehat{\triangledown}_{\mu} \textbf{X}_{\nu} -  \widehat{\triangledown}_{\nu} \textbf{X}_{\mu}
+ g \textbf{X}_{\mu} \times \textbf{X}_{\nu} , \label{eq41}
\end{equation}
with
\begin{eqnarray}
\widehat{\textbf{F}}_{\mu\nu} &=& [F^{V}_{\mu\nu} +  (\partial_{\mu} A^{3}_{\nu} - \partial_{\nu} A^{3}_{\mu})]  \textbf{n}   \nonumber\\
\textbf{X}_{\mu} &=& A^{1}_{\mu} \textbf{n}_{1} + A^{2}_{\mu} \textbf{n}_{2} \nonumber\\
 \widehat{\triangledown}_{\mu} \textbf{X}_{\nu} &=&  \partial_{\mu} \textbf{X}_{\nu} + g (\textbf{A}^{V}_{\mu} + A^{3}_{\mu} \textbf{n} ) \times \textbf{X}_{\nu}   . \label{eq42}
\end{eqnarray}
Lagrangian (\ref{eq40}) is the same as (\ref{eq039}) if we put $ A_{\mu} = A^{V}_{\mu} + A^{3}_{\mu}  $.
Apparently, In the final Lagrangian $ \textbf{n}_{1} $ and $ \textbf{n}_{2} $ are also present, but this could not be a problem, because Lagrangian (\ref{eq40}), as well as vacuum Lagrangian $ L_{V} $, is invariant under (infinitesimal) rotation of internal basis:
\begin{equation}
\delta \textbf{n}_{1} = - \textbf{a} \times \textbf{n}_{1}, \, \, \delta \textbf{n}_{2} = - \textbf{a} \times \textbf{n}_{2}, \, \, \delta \textbf{n} = - \textbf{a} \times \textbf{n},  \label{eq43}
\end{equation}
and
\begin{equation}
\delta A^{V}_{\mu} = \frac{1}{g} \textbf{n} . \partial _{\mu} \textbf{a} ,  \label{eq44}
\end{equation}
hence, one can eliminate $ \textbf{n}_{1} $ and $ \textbf{n}_{2} $ by choosing a gauge, unitary gauge, in which $ \textbf{n} $ and $ \textbf{A}^{V}_{\mu} $ have only third component in the internal space.
Note that Lagrangian (\ref{eq40}) is invariant under another transformation: $  A^{V}_{\mu} \rightarrow A^{V}_{\mu} + A_{\mu} $ and $  A^{3}_{\mu} \rightarrow A^{3}_{\mu} - A_{\mu} $ which $ A_{\mu} $ is an arbitrary four-vector.
In fact, the interaction of $ A^{V}_{\mu} $ and $ A^{3}_{\mu} $ with other fields are the same and they are not recognizable from each other in the full Lagrangian, therefore it is even possible to eliminate $ A^{V}_{\mu} $ by choosing $ A_{\mu} = - A^{V}_{\mu}  $.
 In this case, in addition to particle degrees of freedom, only dimensionless vacuum field $ \textbf{n} $ are present and this background field can interact with quantum fields or particles. We again reached to the same result as the previous section.

\section{Conclusion} \label{sec5}

A novel approach for spontaneous symmetry breaking has been presented. We show that Yang-Mill-Higgs theory based on new variables, which do not change the dynamic of the theory at least at the classical level, leads to different result for symmetry breaking in the non-Abelian case.
In this case apart from particle degrees of freedom, quantum fields, topological degrees of freedom as classical fields also emerge. We call these extra fields
"vacuum fields".
Although it is possible to quantize these fields, but if we treat them like classical fields, then they do not change the particle degrees of freedom after symmetry breaking. For example, in the non-Abelian case, before breaking of SU(2) symmetry we had 3 massive scalar fields $ \bm{\phi} $ and 3 massless vector fields $ \textbf{A}_{\mu} $ with $ 3 + 3 \times 2 = 9 $ total degrees of freedom, and after symmetry breaking we have 1 massive scalar field $ \phi $, 1 massless vector field $ A^{3}_{\mu} $, and 2 massive vector fields $ A^{1}_{\mu} $ and $ A^{2}_{\mu} $ with $ 1 + 1 \times 2 + 2 \times 3 = 9 $. If we consider vacuum fields as the particles and quantize them, the situation changes and extra degrees of freedom appear in the particle spectrum. Besides, if we treat the vacuum fields as classical fields, and not operators, then commutators relations in quantum theory before and after symmetry breaking do not change: for $ \phi_{i} \rightarrow \phi^{V}_{i} + \phi^{phys}_{i}  $ we have $ [\phi_{i} , \phi_{j}] = [\phi^{phys}_{i} , \phi^{phys}_{j}] $ because $ [\phi^{V}_{i} , \phi^{V}_{j}] = 0 $.
Note that our vacuum fields $ \alpha $, and $ \beta $ are massless fields and the vacuum sector of the theory is scale invariant. These fields, unlike any fields in particle physics, are dimensionless fields. We can call these vacuum fields unparticle stuff, though the only unparticle stuff that has been known in physics are classical fields or classical solutions of the field equations known as topological fields.
We emphasize that our vacuum fields appeared because we simply do not consider one vacuum state for all space-time points and the vacuum state itself is a function of space-time: $ \bm{\phi}^{V}(x) =   \nu \, \textbf{n}(x) $. This possibility could be true at least on cosmological scales.
On the other hand, Higgs mechanism works very well in particle physics, so the vacuum fields should be very little on elementary particle space-time scales, but, they could have cosmological consequences in large scales. Indeed, there are theories for both inflation and dark energy which scalar fields, take a crucial role. In our approach, massless scalar fields arise which seems better candidate for inflation and dark energy than massive Higgs field.
In a subsequent paper \cite{Mohamadnejad:2017bvr} we will apply our approach to the more realistic case, Standard Model of particle physics, and we study vacuum sector of this model and discuss its cosmological consequences.

\begin{small}

\providecommand{\href}[2]{#2}
\end{small}

\end{document}